\documentclass[conference]{IEEEtran}
 \usepackage{fancyhdr} 
\ifCLASSINFOpdf
  \usepackage[pdftex]{graphicx}
\else
  \usepackage[dvips]{graphicx}
\fi
\usepackage[cmex10]{amsmath}
\usepackage{url}

\usepackage[final]{microtype}
\microtypesetup{%
  expansion=false,
  protrusion=true,
  verbose=errors,
}

\usepackage{listings}

\usepackage{xspace}

\usepackage{placeins}
\usepackage{fancyvrb}
\usepackage{cite}

\usepackage{algorithm}
\usepackage[noend]{algpseudocode}
\algrenewcommand\algorithmicindent{0.15em}%
\makeatletter
\algrenewcommand\ALG@beginalgorithmic{\small}
\algrenewcommand\algorithmiccomment[2][\normalsize]{{#1\hfill\(\triangleright\) #2}}
\makeatother
\newcommand{\pluseq}{\mathrel{+}=}
\usepackage{graphicx}
\usepackage{caption}
\usepackage{subcaption}
\usepackage{tabularx}
\usepackage{enumitem}   
\graphicspath{ {./figures/} }
\renewcommand{\footnotesize}{\scriptsize}
\makeatletter
\algrenewcommand\ALG@beginalgorithmic{\footnotesize}
\makeatother

\usepackage{color, soul} 
\soulregister\emph7
\soulregister\cite7
\soulregister\ref7

\begin{document}
\setlength{\abovedisplayskip}{3pt}
\setlength{\belowdisplayskip}{3pt}

\setlength\floatsep{0.3\baselineskip plus 1pt minus 20pt}
\setlength\textfloatsep{0.3\baselineskip plus 1pt minus 20pt}
\setlength\intextsep{0.3\baselineskip plus 1pt minus 20pt}

%
\title{Anatomy Of High-Performance Deep Learning Convolutions On SIMD Architectures}

\author{\IEEEauthorblockN{Evangelos Georganas, Sasikanth Avancha, Kunal Banerjee, Dhiraj Kalamkar,\\ Greg Henry, Hans Pabst, and Alexander Heinecke}
\IEEEauthorblockA{Intel Corporation}
\vspace{-6ex}

}

\maketitle

\begin{abstract}
Convolution layers are prevalent in many classes of deep neural networks, including Convolutional Neural Networks (CNNs) which provide state-of-the-art results for tasks like image recognition, neural machine translation and speech recognition. The computationally expensive nature of a convolution operation has led to the proliferation of implementations including matrix-matrix multiplication formulation, and direct convolution primarily targeting GPUs. In this paper, we introduce direct convolution kernels for x86 architectures, in particular for Xeon and Xeon\,Phi systems, which are implemented via a dynamic compilation approach. Our JIT-based implementation shows close to theoretical peak performance, depending on the setting and the CPU architecture at hand. We additionally demonstrate how these JIT-optimized kernels can be integrated into a light-weight multi-node graph execution model. This illustrates that single- and multi-node runs yield high efficiencies and high image-throughputs  when executing state-of-the-art image recognition tasks on CPUs.
\end{abstract}
%
\IEEEpeerreviewmaketitle

\section{Introduction and Related Work}
\label{sec:intro}

In the last few years, deep learning has evolved into one of the most
important computational concepts. Several academic groups and companies have
released open source frameworks which abstract many implementation details from
the data scientist: Tensorflow~\cite{tensorflow2015-whitepaper}, Caffe~\cite{jia2014caffe},
to mention the most popular ones according to GitHub
stars. Additionally, hardware vendors started to provide custom silicon for 
deep learning training, such as the NVidia V100, the Intel Knights Mill processor
and Google's TPU accelerator.  

Although these different frameworks may emphasize distinct workloads,
one of the most important application scenario of neural networks is image
recognition~\cite{origalexnet}. This is implemented via so-called
convolutional neural nets (CNN), e.g.~\cite{convnet-benchmarks}. Layers of
widely-used network topologies are based on small convolutions which can be
easily mapped onto the aforementioned CPUs and GPUs via library functions. 

Achieving close to peak performance in these libraries is essential as most of the 
application execution time is spent here. Often this is done by
flattening corresponding input data (im2col operations) and calling a
standard matrix multiplication (GEMM) afterwards as described in~\cite{cudnn,vasudevan2017parallel,anderson2017low}. However, two
downsides can be seen for this approach: one is the memory footprint overhead and the other is the
introduction of a memory bandwidth dependency in a computationally expensive
operation.  The latter downside might create a huge performance penalty on CPU
architectures. Therefore, a new flavor of implementation has started
to emerge recently, called {\em direct convolution}. In this approach, a convolution is
directly applied to the layers of the CNN.  By leveraging this strategy, we avoid
costly memory operations such as vector shuffle, gather, and/or scatter. Other layers such
as ReLU, Pooling, LRN, Normalization, Batch-concatenation do not impose any
memory layout requirements.  These layers can be efficiently
implemented on any layout which maximizes the performance benefit of
convolutional layers. 

As mentioned before, a huge fraction of the workload, especially when training a neural network,
is spent in GEMM-flavored compute or convolution operations. This nominates deep learning training as 
one the most important next generation HPC scale-out application candidates. Recently several research groups have showcased how the training task
can be scaled to a large number of nodes and clusters with multiple TFLOPS to PFLOPS of compute~\cite{goyal2017accurate,cho2017powerai,pcldnn}.
There is a rich research landscape in regard to parallelizing DNN training as
summed up in~\cite{bennun2018demyst}; the best approach to reduce the overall time-to-train is to aim for the fastest
single node performance and to scale this performance out. Achieving the best 
possible single node performance on CPUs is one of the major contributions of this work.

For direct convolutions, meta-programming via templates (e.g.~\cite{ZNN}) or
static compilation (e.g.~\cite{neon}) are often employed to achieve close to
peak performance on a given architecture. This approach not only imposes a
static compilation step, but also requires fine-tuning for each topology
separately.  
Additionally, prior work~\cite{Heinecke:2016:LAS:3014904.3015017} has shown that statically-tuned BLAS-calls incur
overheads for small GEMMs and therefore do not achieve the highest performance
on x86 systems.  It is proposed to use runtime code specialization via JIT-ing for small
GEMMs and achieve close to peak performance.  Since the matrices involved in
convolutional neural networks are typically tall and skinny, we employ a similar JIT-ing
strategy to implement fast direct convolutions on CPUs in this paper.  We lay
out the convolution's tensor data for input, output and filter in a vectorization- and
cache-friendly manner, and apply standard compiler optimizations such as
register and cache blocking, which are theoretically analyzed in~\cite{demmel2018communication}.  Some of the key optimizations we apply include
software prefetching, per-thread-based access optimizations and layer fusion 
within the CNN topology. The goal is to reduce passes over the data to 
an absolute minimum. However, layer fusion results in growing significantly the number of required kernels as many different combinations of fused layer patterns are required. This is yet another
reason to move away from static compilation towards a runtime and on-demand driven
compiling infrastructure.  One thing to keep in mind is that our JIT does not incur the
overheads of recompilation and tuning. 


The main contributions of this paper are:
\begin{itemize}
  \item deriving and defining the ingredients of fast direct convolution kernels for training CNNs on modern CPU architectures.
  \item showcasing how JIT compilation and a layer/execution graph  strategy can be combined to master the combinatorial explosion in the number of required kernels and to increase data locality. This includes layer fusion which today is not available in vendor's libraries.
  \item a careful performance study of various and most recent CPU architectures on a kernel- and multi-node level for CNN training.
\end{itemize}


\section{Implementation}
\label{sec:impl}
Before diving into the specifics of our implementation, we introduce some basic terminology and notation. A neural network consists of layers of multiple neurons connected by weights. The values assigned to a neuron are usually called activations. Both activations and weights are represented with multidimensional tensors. Any activation tensor can be further categorized as \emph{input} or \emph{output}. The activation tensors conceptually consist of 4 dimensions: the minibatch size $N$, the number of feature maps $C$ and the spatial dimensions $H$ and $W$. Throughout this paper, we denote the input tensor dimensions with $N$, $C$, $H$ and $W$ while the corresponding output tensor dimensions are $N$, $K$ (output feature maps), $P$ and $Q$ (output spatial dimensions). The weight tensor is conceptually characterized also by 4 dimensions: the feature map dimensions $C$, $K$ and the spatial dimensions $R$ and $S$.
\subsection{Forward propagation loop structure}
\begin{algorithm}[t]
\begin{algorithmic}[1]
\For{$n=0 \dots N-1$}
\For{$k=0 \dots K-1$}
\For{$c=0 \dots C-1$}
\For{$oj=0 \dots P-1$}
\For{$oi=0 \dots Q-1$}
\State $ij = stride * oj$
\State $ii = stride * oi$
\For{$r=0 \dots R-1$}
\For{$s=0 \dots S-1$}
\State $O[n][k][oj][oi]\pluseq I[n][c][ij+r][ii+s]*W[k][c][r][s]$
\EndFor
\EndFor
\EndFor
\EndFor
\EndFor
\EndFor
\EndFor
\end{algorithmic}
\caption{Naive forward propagation loops}
\label{alg:naive_fwd}
\end{algorithm}
The forward propagation layer consists of seven nested loops that convolve the input tensor $I$ and the weight tensor $W$, yielding the output tensor $O$ (see Algorithm~\ref{alg:naive_fwd} that implements the direct convolution method). The input spatial domain may be accessed in a strided way, dictated by the parameter $stride$. In the following subsections we incrementally introduce the optimizations of the nested loops.

\subsection{Vectorization and register blocking}
\label{subsec:vec}
\begin{algorithm}[t]
\begin{algorithmic}[1]
\State $C_b = C/VLEN$
\State $K_b = K/VLEN$
\State $P_b = P/RB_P$
\State $Q_b = Q/RB_Q$
\For{$n=0 \dots N-1$}
\For{$k_b=0 \dots K_b-1$}
\For{$c_b=0 \dots C_b-1$}
\For{$ojb=0 \dots P_b-1$}
\For{$oib=0 \dots Q_b-1$}
\State $ij = stride * ojb * RB_P$
\State $ii = stride * oib * RB_Q $
\State $oj = ojb * RB_P$
\State $oi = oib * RB_Q$
\For{$r=0 \dots R-1$}
\For{$s=0 \dots S-1$}
\For{$k=0 \dots VLEN$}
\For{$c=0 \dots VLEN$}
\For{$p=0 \dots RB_P$}
\For{$q=0 \dots RB_Q$}
\State $ij^\prime = ij + stride * p$
\State $ii^\prime = ii + stride * q$
\State $\thinmuskip=0mu \medmuskip=0mu \thickmuskip=0mu  O[n][k_b][oj+p][oi+q][k]\pluseq W[k_b][c_b][r][s][c][k]*$
\State $\qquad \qquad \qquad \qquad \qquad \qquad  I[n][c_b][ij^\prime+r][ii^\prime+s][c]$
\EndFor
\EndFor
\EndFor
\EndFor
\EndFor
\EndFor
\EndFor
\EndFor
\EndFor
\EndFor
\EndFor
\end{algorithmic}
\caption{Forward propagation with register blocking}
\label{alg:rb_fwd}
\end{algorithm}
We observe that the output feature maps can be computed independently in a data-parallel fashion. Thus, in order to vectorize the fused multiply-add (FMA) operation at line 10 of Algorithm~\ref{alg:naive_fwd} we opt to block the feature maps by a factor of $VLEN$ and we pull the vectorization block as the innermost, fast-running dimension of the tensors. $VLEN$ is a parameter which depends on the vector register width of the target architecture and the tensor datatype. For instance, given an AVX512 architecture and FP32 tensor datatype, $VLEN$ is 16. In addition to the vectorization for the feature map dimensions, register blocking is used to improve data reuse from registers, decrease L1 cache traffic, and most importantly to hide the latency of the FMA instructions. We apply register blocking in the spatial domains of the output tensor since points in the spatial iteration space can be computed independently. In this way, we form independent accumulation chains in registers that are sufficient to hide FMA latencies. Algorithm~\ref{alg:rb_fwd} illustrates the convolution loops rewritten in a way that exposes the register blocking and the vectorization opportunities. The register blocking factors $RB_P$ and $RB_Q$ are chosen based on the architectural target and are further discussed in subsection~\ref{subsec:micro}.

\subsection{Cache blocking and loop ordering}
\label{subsec:cache_block}
Unless the activations and the weight tensors fit in cache, the convolution loops of Algorithm~\ref{alg:rb_fwd} can be bandwidth bound. To maximize data reuse from cache, we also apply cache blocking in the feature map and spatial dimensions of Algorithm~\ref{alg:rb_fwd}. Also, the loop ordering determines the way the tensors are accessed and impacts the reuse of the corresponding data~\cite{demmel2018communication}. For large values of weight spatial domains e.g.\ $R=3$, $S=3$ and given the loop ordering of Algorithm~\ref{alg:rb_fwd}, the output tensor entries can be reused from registers multiple times. On the contrary, for convolution layers with $R=1$, $S=1$, the output tensor entries do not employ the same degree of register reuse. However, if we pull in the input feature map loop (line 7 of Algorithm~\ref{alg:rb_fwd}), then we can increase the register reuse for the output tensor by a factor of $C_b$.

\subsection{Code generation for convolution microkernel}
\label{subsec:micro}
\begin{algorithm}[t]
\begin{algorithmic}[1]
\For{$n=0 \dots N-1$}
\For{$k_b=0 \dots K_b-1$}
\For{$c_b=0 \dots C_b-1$}
\For{$ojb=0 \dots P_b-1$}
\For{$oib=0 \dots Q_b-1$}
\State $ij = stride * ojb * RB_P$
\State $ii = stride * oib * RB_Q $
\State $oj = ojb * RB_P$
\State $oi = oib * RB_Q $
\State $\thinmuskip=0mu \medmuskip=0mu \thickmuskip=0mu CONV(\&I[n][c_b][ij][ii][0], \&W[k_b][c_b][0][0][0][0], \thinmuskip=0mu \medmuskip=0mu \thickmuskip=0mu \&O[n][k_b][oj][oi][0])$
\EndFor
\EndFor
\EndFor
\EndFor
\EndFor
\end{algorithmic}
\caption{Forward propagation with microkernel calls}
\label{alg:micro_fwd}
\end{algorithm}
In Algorithm~\ref{alg:rb_fwd}, the loops in lines 14 - 23 are written as a JIT-ed high performance microkernel that performs a small convolution. This microkernel takes essentially three arguments: a pointer to the output sub-tensor that is computed by the kernel invocation and the corresponding pointers of the required input and weight sub-tensors. Algorithm~\ref{alg:micro_fwd} illustrates the forward propagation algorithm implemented with such a convolution microkernel.

It can be seen in Algorithm~\ref{alg:rb_fwd} that the inner-most computation is a small matrix-vector product of a partial weight tensor with a partial input tensor: $O'[k] \pluseq W'[c][k] * I'[c]$. The $c,k$ dimensions are multiples of the architecture's vector lengeth $VLEN$ and therefore have normally the values 16. 
A matrix-vector product is not compute intense, however we can see that there is a lot of reuse of 
either the output or the weight tensor data when taking the outer loops into account. This turns the small matrix-vector product into a sequence
of small matrix multiplications (GEMM) with a blocking in $RB_Q$. As a simple introductory 
example, let us choose following convolution size parameters: $R=S=1$ and $RB_P=1$. In this case the linear algebra expert eye
realizes a matrix multiplication with the following dimensions in BLAS notation dimensions: $\hat{M}=k$, $\hat{N}=RB_Q$, $\hat{K}=c$. As the value
$RB_Q$ heavily depends on the convolution at hand, we unfortunately can not hard-code an one-fits-all GEMM kernel. Instead, we implemented a runtime
just-in-time (JIT) code generator following the ideas presented in~\cite{Heinecke:2016:LAS:3014904.3015017}, while optimizing for $\hat{M}$ and $\hat{K}$ being multiples of the machine's 
vector length. In this case it is important that $\hat{N}=RB_Q$ is larger than the machine's FMA latency (see above Section~\ref{subsec:vec}). 
As we target Intel AVX512 enabled platforms in this work (see Section~\ref{sec:performance} for a detailed hardware list) the small GEMM kernel uses the following basic block: a) loading 
a full vector-register with output channels weights from $W$ at position $0 \leq x < c$ and b) loop over $RB_Q$ pixels of the input activation, broadcasting those and 
multiplying them with the loaded weights. This results into following JIT'ed GEMM code: for $0 \leq x < c$ and $0 \leq y < RB_Q$ do $O'[y][k] \pluseq W'[x][k] * I'[y][x]$.

However, just having a small GEMM kernel JIT'ed is not enough to achieve sufficient performance in an arbitrary convolutional layer of a CNN. Two additional
optimizations are needed: a) load/store optimization of $O$ in case of $R,S>1$ b) additional pixel blocking when $Q=RB_Q$ and this value is smaller than FMA latency.
The solution for a) is straight-forward. In this case we run a sequence of small GEMMs which write to the same result matrix: $O'[v][w][:] = \sum_{r=0,s=0}^{r=R,s=S}
W'[r][s][:][:] * I'[v+r][w+s][:]$ and we hoist the writes to $O$ outside of the $R,S$ loops. In case of b) we run two small GEMMs in the same JIT'ed kernel which share the same
weight matrix:  $O'[t][:][:] = W'[:][:] * I'[t][:][:]$ with $0<t<RB_P$. Here $RB_Q * RB_P$ should be larger than the machine's FMA latency. Of course both concepts can be combined. These two optimizations also highlight the benefits of a specialized convolution kernel and a batched GEMM approach for the independent small GEMMs (not reducing into one C-matrix) is not able to leverage these two concepts.
\subsection{Prefetching}
\label{subsec:prefetching}
An important optimization in modern CPU architectures is software prefetching that aims to mitigate cache miss latency overheads. The microkernel described in subsection~\ref{subsec:micro} is further enriched with prefetching capabilities. More specifically, software prefetch instructions are sprinkled throughout the FMA instructions and effectively prefetch sub-tensors to be used by future FMA instructions. In our implementation we design a two-level prefetch strategy. At the first level, we issue L1 cache prefetches pulling in data to be used ``later" (within a tunable temporal distance) \emph{by the same microkernel invocation}. At the second level, we issue L2 cache prefetches involving sub-tensors of \emph{future microkernel invocations}. In order to accommodate the second level of prefetching, we extend the microkernel API with three additional arguments: a pointer to an output sub-tensor that will be used by a future invocation and the pointers of the required input and weight sub-tensors to be prefetched.

Such a two-level prefetch strategy virtually diminishes cache miss latency overheads from the critical path. However, finding the correct pointers of sub-tensors that will be used in future microkernel invocations and using them as convolution arguments requires a complicated, branchy logic. This  branchy logic assesses the boundaries of the five-dimensional iteration space (lines 1 - 5 in Algorithm~\ref{alg:micro_fwd}) and calculates the proper sub-tensor offsets of future kernel invocations. 

\subsection{Parallelization strategy}
In principle, there is abundant parallelism available since loops at lines 1, 2, 4 and 5 define output tensor slices that can be processed independently. More accurately, there are $N \times K_b \times P_b \times Q_b$ independent microkernel invocations (or equivalently ``work items") that can be assigned to the available threads. First, we opt to divide work based on the minibatch iteration (line 1 of Algorithm~\ref{alg:micro_fwd}); in this way, threads share the entire weights tensor which subsequently can be reused from shared caches. In case the minibatch domain does not provide sufficient parallelism, we further extract work items from the output feature map domain. Finally, if the number of threads is greater than the $N\times K_b$ work items, we further utilize parallelism from the spatial domains $P_b$ and $Q_b$ of the output tensor.


\subsection{Layer fusion}
\begin{algorithm}[t]
\begin{algorithmic}[1]
\For{$n=0 \dots N-1$}
\For{$k_b=0 \dots K_b-1$}
\For{$c_b=0 \dots C_b-1$}
\For{$ojb=0 \dots P_b-1$}
\For{$oib=0 \dots Q_b-1$}
\State $ij = stride * ojb * RB_P$
\State $ii = stride * oib * RB_Q $
\State $oj = ojb * RB_P$
\State $oi = oib * RB_Q $
\State $\thinmuskip=0mu \medmuskip=0mu \thickmuskip=0mu CONV(\&I[n][c_b][ij][ii][0], \&W[k_b][c_b][0][0][0][0],\&O[n][k_b][oj][oi][0])$
\If{$\textit{f}use(L())\ \ \textbf{AND}\ \  c_b == C_b-1$}
\State $APPLY(L(),\ \&O[n][k_b][oj][oi][0])) $
\EndIf
\EndFor
\EndFor
\EndFor
\EndFor
\EndFor
\end{algorithmic}
\caption{Forward propagation with fused layer L()}
\label{alg:fusion_fwd}
\end{algorithm}

Modern DNN architectures consist not only of convolution layers, but they also contain layers like ReLU, Pooling, LRN, Normalization and Bias. Some of these layers can be materialized by applying a function $L()$ to a tensor and such non-convolution layers typically have low operational intensity, hence they are bandwidth bound. In our framework we identify and exploit layer fusion opportunities, i.e.\ we decompose these various operational layers such that they operate on sub-tensors and we apply them when the involved data are hot in cache, e.g.\ due to convolution. By taking advantage of such temporal locality, we save memory bandwidth that these layers would otherwise consume. In Algorithm~\ref{alg:fusion_fwd} we illustrate an example, where we fuse in the forward propagation an operator $L()$ after an output sub-tensor has been fully computed and is hot in cache. We observe in Algorithm~\ref{alg:fusion_fwd} that the layer fusion requires conditional statements to determine when to apply the relevant operator. 

\subsection{Kernel streams}
\label{subsec:KS}
\begin{figure*}[t]
\centering
\includegraphics[width=1.9\columnwidth]{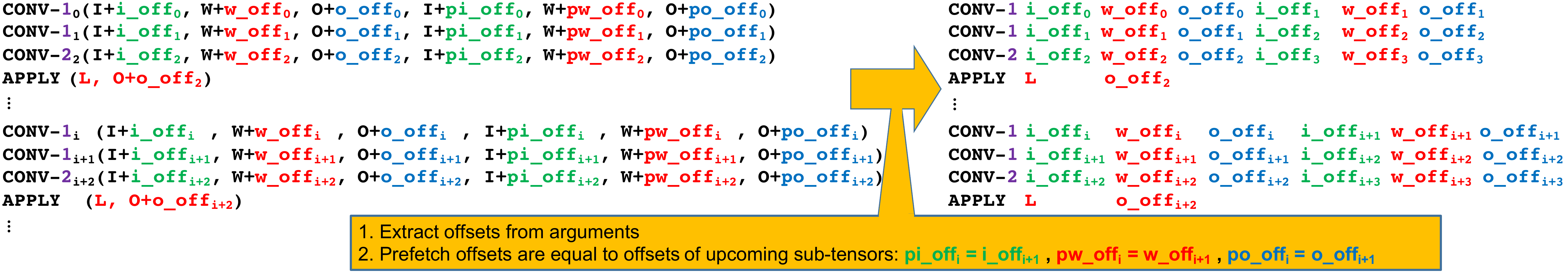}
\caption{Stream of calls during the execution of convolution loops}
\vspace{-15pt}
\label{fig:calls}
\end{figure*}

As described in the previous subsections, finding the optimal prefetching arguments for the microkernel invocations and enabling layer fusion introduces complicated, conditional code segments in the main loops that incur overhead at runtime. Additionally, the way Algorithm~\ref{alg:micro_fwd} is written implies that only one variant of convolution microkernel is required. Even though in general this is true, there are cases where the spatial dimensions $P$ and $Q$ are not perfectly divided by the register blocking factors $RB_P$ and $RB_Q$. In such a scenario, instead of sacrificing performance by reducing the size of the register blocking factors, we can generate a second microkernel with register blocking factors $RB^\prime_P$ and $RB^\prime_Q$. The latter convolution kernel should be executed at the boundaries of the loops controlling the spatial dimensions $P$ and $Q$ (lines 4 and 5 of Algorithm~\ref{alg:micro_fwd}). Therefore, finding which microkernel variant to execute at every iteration requires yet another conditional statement.

We address all these issues by introducing a framework called \emph{kernel streams} consisting of two phases: the \emph{dryrun} and the \emph{replay} phase. The kernel streams framework is inspired by the following key observation: During the execution of the convolution loops, each thread performs a series of calls to the convolution microkernels which may be interleaved with other kernels/operators in case of layer fusion. For example, in the left part of Figure~\ref{fig:calls} we illustrate the series of calls a thread performs during runtime. We observe that there are  two types of calls: Calls to the convolution microkernels and calls to other operators due to fusion.

The convolution microkernels may have multiple variants (e.g.\ CONV-1, CONV-2) depending on the register blocking factors that each variant is using. Furthermore, each convolution kernel takes six arguments after the enabling of prefetching described in subsection~\ref{subsec:prefetching}. The first three arguments are pointers to the input, weight and output sub-tensors involved in the computation of the current convolution, while the last three arguments are pointers to the input, weight and output sub-tensors that will be prefetched throughout the kernel execution. Even though each argument is effectively a pointer to a sub-tensor, it can be represented as an offset added to the base pointer of the corresponding tensor, and this is how the convolution kernel calls are written in Figure~\ref{fig:calls}. We further make the following observation: The prefetch offsets $pi\_o\mathit{ff}_i$, $pw\_o\mathit{ff}_i$ and $po\_o\mathit{ff}_i$ for a convolution at step $i$ should be equal to the offsets $i\_o\mathit{ff}_{i+1}$, $w\_o\mathit{ff}_{i+1}$ and $o\_o\mathit{ff}_{i+1}$ of the sub-tensors consumed in the convolution at step $i+1$. This is the case because at step $i$ we want to prefetch the sub-tensors to be used at step $i+1$. In practice, we tune the prefetch distance based on the computational cost of the convolution kernel and the corresponding layer. By using the offsets of the sub-tensors as arguments, and by leveraging the aforementioned property of the prefetch offsets, we can rewrite the stream of convolution kernel calls as they appear in the Right part of Figure~\ref{fig:calls}. Regarding calls to other operators due to fusion, we denote them in the stream of execution as calls to a kernel APPLY followed by the specific function $L()$ that is fused and the proper sub-tensor argument/offset.

\begin{figure}[t]
\centering
\includegraphics[width=1\columnwidth]{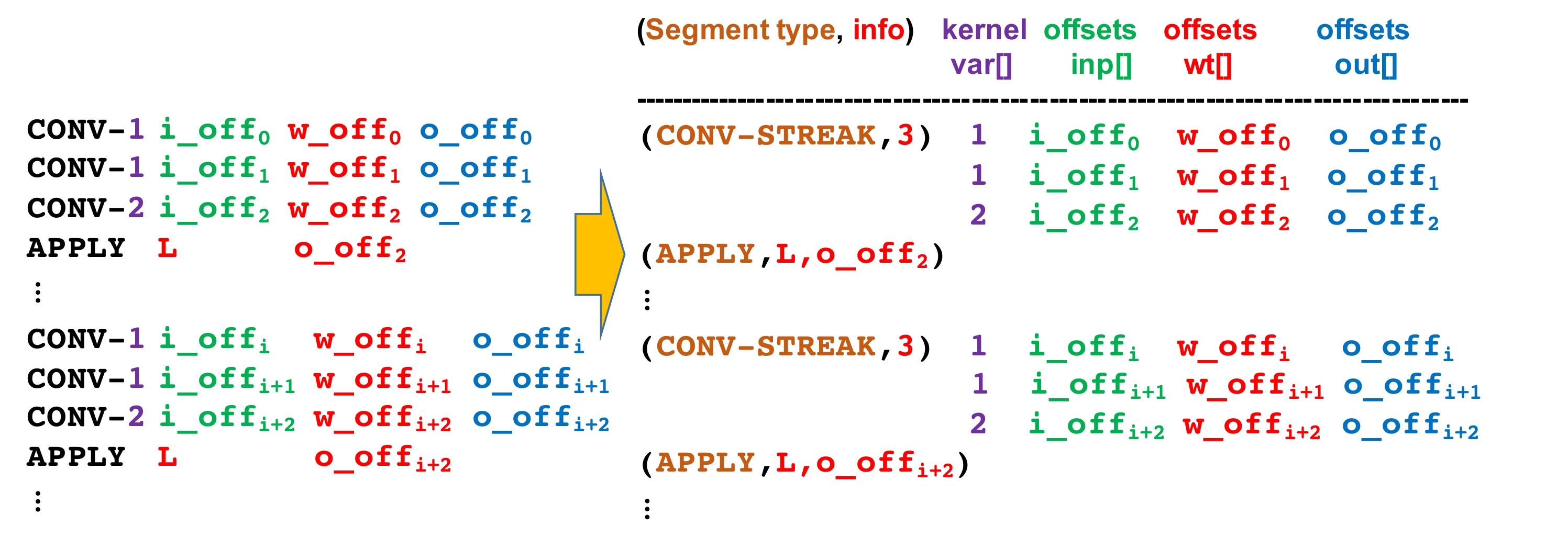}
\vspace{-10pt}
\caption{Stream of calls encoded as segments and offset streams}
\label{fig:streams}
\end{figure}

Given the formulation of Figure~\ref{fig:calls}, in order to perform the forward propagation we need 5 streams, shown in the Left part of Figure~\ref{fig:streams}: i) a stream for the kernel type (CONV-1,CONV-2, or APPLY), ii) a stream of input offsets, iii) a stream of weight offsets, iv) a stream of output offsets and v) a stream of arguments for the APPLY kernels. Typically a sequence of convolution calls is followed by a fused operation, which is then subsequently followed by another streak of convolutions. We take advantage of this structure and we further encode the entire forward propagation as \emph{segments}, representing either streaks of convolutions (CONV-STREAK) or fused operators (APPLY). Along with the stream of convolution kernel variants $var$, and the sub-tensors's offset streams $inp$, $wt$ and $out$ we have a compact representation of the entire forward propagation (see Right part of Figure~\ref{fig:streams}) which can be simply re-written as in Algorithm~\ref{alg:streams_fwd}. Algorithm~\ref{alg:streams_fwd} represents the \emph{replay} phase of the kernel streams framework.

\begin{algorithm}[t]
\begin{algorithmic}[1]
\algrenewcommand\algorithmicindent{1.05em}%
\State $i=0$
\For{$pc=0 \dots n\_segments-1$}
\If{$segment[pc].type == \textit{CONV-STREAK}$}
\State $n\_convs =  segment[pc].in\textit{f}o$
\For{$ci=0 \dots n\_convs-1$}
\State $\thinmuskip=0mu \medmuskip=0mu \thickmuskip=0mu CONV[var[i]](I+inp[i], W+wt[i], O+out[i],$
\State $\thinmuskip=0mu \medmuskip=0mu \thickmuskip=0mu \ \ \ \ \ \ \ \ \ \ \ \ \ \ \ \ \ \ \ \ \ \ \ \ \ \ I+inp[i+1], W+wt[i+1], O+out[i+1])$
\State $\thinmuskip=0mu \medmuskip=0mu \thickmuskip=0mu i+=1$
\EndFor
\EndIf
\If{$segment[pc].type == \textit{APPLY}$}
\State $APPLY(L(), segment[pc].in\textit{f}o) $
\EndIf
\EndFor
\end{algorithmic}
\caption{Forward propagation via replay of kernel streams}
\label{alg:streams_fwd}
\end{algorithm}

We generate the prerequisite arguments (segments and streams) of Algorithm~\ref{alg:streams_fwd} at the \emph{dryrun} phase of our framework. In the dryrun phase, we perform the loops dictated by Algorithm~\ref{alg:fusion_fwd} but instead of making calls to kernels, we record the proper arguments/offsets and the types of the kernel calls in auxiliary \emph{stream buffers}. We emphasize here that the generation of the stream buffers are thread-specific since each thread is assigned a distinct output sub-tensor. We further encode these streams into segments as in Figure~\ref{fig:streams} by leveraging a specialized run-length encoding procedure. Similarly to the JIT-ing of the convolution microkernels, the dryrun phase has to be performed only once during the setup of the CNN layer; during runtime, we perform the replay phase of the kernel streams framework.


\subsection{Backward propagation implementation}
\label{subsec:duality}
\begin{algorithm}[t]
\begin{algorithmic}[1]
\For{$n=0 \dots N-1$}
\For{$k=0 \dots K-1$}
\For{$c=0 \dots C-1$}
\For{$oj=0 \dots P-1$}
\For{$oi=0 \dots Q-1$}
\State $ij = stride * oj$
\State $ii = stride * oi$
\For{$r=0 \dots R-1$}
\For{$s=0 \dots S-1$}
\State $dI[n][c][ij+r][ii+s]\pluseq dO[n][k][oj][oi]*W[k][c][r][s]$
\EndFor
\EndFor
\EndFor
\EndFor
\EndFor
\EndFor
\EndFor
\end{algorithmic}
\caption{Naive backward propagation loops}
\label{alg:naive_bwd}
\end{algorithm}

The back propagation algorithm is described by the loop structure of Algorithm~\ref{alg:naive_bwd}. In the back propagation pass, we compute the gradient input tensor $dI$ by convolving the gradient output tensor $dO$, with the weight tensor $W$. At first sight, this algorithm is different from forward propagation, since the accumulation happens into the gradients of inputs and its eventual update:
\begin{equation*}
dI[n][c][ij+r][ii+s]\pluseq dO[n][k][oj][oi]*W[k][c][r][s]
\end{equation*}
has different access pattern than the update of forward propagation:
\begin{equation*}
O[n][k][oj][oi]\pluseq I[n][c][ij+r][ii+s]*W[k][c][r][s]
\end{equation*}
We show here that in two scenarios (which cover the majority of contemporary CNN layers) we can transform the weight tensors, and then we can reuse the high performance forward propagation described in the previous subsections.

1) \textbf{Scenario with $\boldsymbol{stride=1}$}. In this case we get $ij = oj$, $ii=oi$. By setting $ij+r=IJ$ and $ii+s =II$ we can rewrite the update of the input gradients: 
\begin{equation*}
dI[n][c][IJ][II]\pluseq dO[n][k][IJ-r][II-s]*W[k][c][r][s]
\end{equation*}
By creating a new weight tensor $W^\prime$ with:
\begin{equation*}
W^\prime[c][k][-r][-s] = W[k][c][r][s]
\end{equation*}
and by setting $-r=r^\prime$ and $-s=s^\prime$ we can rewrite the update of the input gradients as:
\begin{equation*}
dI[n][c][IJ][II]\pluseq dO[n][k][IJ+r^\prime][II+s^\prime]*W^\prime[c][k][r^\prime][s^\prime]
\end{equation*}
which matches the access pattern of the forward propagation.

2) \textbf{Scenario with $\boldsymbol{R=1}$ and $\boldsymbol{S=1}$}. In this case we always have $r=0$, $s=0$, so the forward update is:
 \begin{equation*}
O[n][k][oj][oi]\pluseq I[n][c][oj*stride][oi*stride]*W[k][c][0][0]
\end{equation*}
By setting $ij=IJ$ and $ii=II$, $oj=ij/stride=IJ/stride$ and $oi=ii/stride=II/stride$ we can rewrite the update of the input gradients: 
\begin{equation*}
dI[n][c][IJ][II]\pluseq dO[n][k][\frac{IJ}{stride}][\frac{II}{stride}]*W[k][c][0][0]
\end{equation*}
By creating a new weight tensor $W^\prime$ with:
\begin{equation*}
W^\prime[c][k][0][0] = W[k][c][0][0]
\end{equation*}
and by setting $1/stride=s^\prime$ we can rewrite the update of the input gradients as:
\begin{equation*}
dI[n][c][IJ][II]\pluseq dO[n][k][IJ*s^\prime][II*s^\prime]*W^\prime[c][k][0][0]
\end{equation*}
which matches the access pattern of the forward propagation.

Therefore, if the layer's specifications fall into one of the above mentioned scenarios, we transform the weight tensor and we leverage the high performance forward propagation. In the remaining cases, we leverage a generic implementation (see Algorithm~\ref{alg:generic_bwd}) that uses small high performance GEMMs to implement the updates of the input gradient tensor. In this formulation, the gradient input and output tensors use the same data layout as the one described in subsection~\ref{subsec:vec} that is amenable to vectorization. Also, the weight tensor is transformed in such a way that the input and output feature map dimensions are transposed while the spatial dimensions are flipped. In these GEMM calls, we follow the convention $GEMM(A,B,C)$ where $A$ is a $M\times K$ matrix, $B$ is a $K\times N$ matrix and $C\pluseq A\times B$. More specifically, our GEMMs have dimensions: $M=VLEN$, $K=VLEN$ and $N=Q$. A small downside of this method is that loops 2, 8 and 9 can not be embedded in a small GEMM call, as such this approach does not exploit all the available data reuse from registers and generates redundant data movement (loads and stores of output sub-tensors).

\begin{algorithm}[t]
\begin{algorithmic}[1]
\For{$n=0 \dots N-1$}
\For{$k_b=0 \dots K_b-1$}
\For{$c_b=0 \dots C_b-1$}
\For{$oj=0 \dots P-1$}
\State $ij = stride * oj$
\State $oi = 0$
\State $ii = 0$
\For{$r=0 \dots R-1$}
\For{$s=0 \dots S-1$}
\State $GEMM(\&W[c_b][k_b][R-1-r][S-1-s][0][0],$
\State $\ \ \ \ \ \ \ \ \ \ \ \  \&dO[n][k_b][oj][oi][0], \&dI[n][c_b][ij+r][ii+s][0])$
\EndFor
\EndFor
\EndFor
\EndFor
\EndFor
\EndFor
\end{algorithmic}
\caption{Backward propagation with small GEMM calls}
\label{alg:generic_bwd}
\end{algorithm}

\subsection{Weight gradient update implementation}
\label{subsec:wup}
\begin{algorithm}[t]
\begin{algorithmic}[1]
\For{$n=0 \dots N-1$}
\For{$k=0 \dots K-1$}
\For{$c=0 \dots C-1$}
\For{$oj=0 \dots P-1$}
\For{$oi=0 \dots Q-1$}
\State $ij = stride * oj$
\State $ii = stride * oi$
\For{$r=0 \dots R-1$}
\For{$s=0 \dots S-1$}
\State $dW[k][c][r][s]\pluseq I[n][c][ij+r][ii+s]*dO[n][k][oj][oi]$
\EndFor
\EndFor
\EndFor
\EndFor
\EndFor
\EndFor
\EndFor
\end{algorithmic}
\caption{Naive weight gradient update loops}
\label{alg:naive_upd}
\end{algorithm}

In the update pass of the weight gradients shown in Algorithm~\ref{alg:naive_upd}, the weight gradient tensor $dW$ is computed by convolving the gradient output tensor $dO$ with the input tensor $I$. By leveraging the same layout for the tensors that is amenable to vectorization and by applying blocking in the spatial dimensions of the $dO$ and $I$ tensors we get the optimized Algorithm~\ref{alg:optimized_upd}. In this optimized loop structure, the last 4 loops (lines 14-21) can be implemented as a JIT-ed microkernel, similar to the one described in subsection~\ref{subsec:micro}. The main difference here is that each microkernel invocation computes a $VLEN\times VLEN$ sub-tensor of the weight gradient. Therefore, we can employ a register blocking up to a factor of $VLEN$ (or equivalently expose $VLEN$ independent FMA instructions). Also, the blocking of the spatial domain with factors $B_P$ and $B_Q$ determines the footprint of the microkernel. By setting $B_P=P$ and $B_Q=Q$ we can maximize the reuse of a $VLEN\times VLEN$ weight gradient sub-tensor/block in registers, however we have to read $H*W*VLEN$ entries of the input tensor and $P*Q*VLEN$ entries of the output gradient tensor. For large spatial dimensions, such a strategy may spill the cache and we will not be able to reuse the input sub-tensor and the output gradient sub-tensor from cache during subsequent kernel invocations. Therefore we opt to block the spatial dimensions depending on the layer characteristics.

\begin{algorithm}[t]
\begin{algorithmic}[1]
\State $P_b = P/B_P$
\State $Q_b = Q/B_Q$
\For{$n=0 \dots N-1$}
\For{$k_b=0 \dots K_b-1$}
\For{$c_b=0 \dots C_b-1$}
\For{$ojb=0 \dots P_b-1$}
\For{$oib=0 \dots Q_b-1$}
\State $ij = stride * ojb * B_P$
\State $ii = stride * oib * B_Q $
\State $oj = ojb * B_P$
\State $oi = oib * B_Q $
\For{$r=0 \dots R-1$}
\For{$s=0 \dots S-1$}
\For{$p=0 \dots B_P$}
\For{$q=0 \dots B_Q$}
\For{$k=0 \dots VLEN$}
\For{$c=0 \dots VLEN$}
\State $ij \pluseq stride * p$
\State $ii \pluseq stride * q$
\State $\thinmuskip=0mu \medmuskip=0mu \thickmuskip=0mu  dW[k_b][c_b][r][s][c][k]\pluseq I[n][c_b][ij+r][ii+s][c]*$
\State $\qquad \qquad \qquad \qquad \qquad dO[n][k_b][oj+p][oi+q][k]$
\EndFor
\EndFor
\EndFor
\EndFor
\EndFor
\EndFor
\EndFor
\EndFor
\EndFor
\EndFor
\EndFor
\end{algorithmic}
\caption{Optimized weight gradient update propagation}
\label{alg:optimized_upd}
\end{algorithm}

In the weight gradient update pass, we have $R\times S \times K_b \times C_b$ independent tasks. If this amount of parallelism is sufficient for $T$ threads and assuming perfect work distribution, then each thread computes $(R*S*C*K)/T$ entries of the weight gradient tensor. Assuming that each thread is assigned $C/T_c$ and $K/T_k$ distinct feature maps, then this parallelization approach requires for each thread to read $(N*C*H*W)/T_c$ input tensor entries and $(N*K*P*Q)/T_k$ gradient output tensor entries.

On the contrary, if we opt for a different parallelization strategy, where each thread computes its own partial, local copy of gradient weights by distributing the minibatch dimension $N$, then we can extract more parallelism (assuming $N>T$). At the end of such an algorithm, the threads have to perform a sum reduction of the $T$ partial weight gradient local copies in order to compute the final weight gradient tensor. In such an approach, each thread computes a partial local copy of the gradient weights with size $R*S*C*K$  and also each thread is required to read $(N*C*H*W)/T$ input tensor entries and $(N*K*P*Q)/T$ gradient output tensor entries. For the final reduction, where each thread is assigned to reduce $(1/T)$-th of the weight gradient tensor copies, each thread has to read in total $R*S*C*K$ weight gradient tensor entries.

The number of operations/computational cost for both parallelization approaches is the same. However, the bandwidth requirements can vary significantly, depending on the layer specifications. More specifically, the first approach requires to read $T/T_c\times$ more input tensor entries and $T/T_k\times$ more gradient output tensor entries compared to the second parallelization approach. However, the latter approach requires to read/write $2T\times$ more weight gradient tensor entries than the first approach. Of course these two parallel algorithms represent two extreme cases: one that uses a single weight gradient tensor and one that utilizes $T$ additional weight gradient tensor copies. We can devise hybrid versions of these two extremes, where we can adjust the number of weight gradient tensor copies by modifying the parallelism over the minibatch dimension. These hybrid algorithms balance the bandwidth requirements of reading the input/gradient output tensors with the bandwidth requirements of reading/writing the gradient weight tensor. Therefore, during the dryrun phase of the weight gradient update propagation we decide on which parallelization strategy to use given the available number of threads and the layer specifications.



\subsection{Reduced Precision: Quantized 16bit Kernels}
 \label{subsec:lp}
Another big trend in deep neural net training is reduced precision to speed-up time-to-train. There are several different solutions available today, whereas GPUs prefer FP16~\cite{micikevicius2017mixed}, CPUs provide an increased throughput for int16 datatypes on the Knights Mill processor through 4VNNIW extensions. The 4VNNIW instruction takes int16 inputs and multiplies and accumulates into int32 values. All of the techniques presented above have been included in kernels which leverage these type of instructions. A proof-of-concept implementation using 4VNNIW has been proven to converge ResNet-50~\cite{he2016deep} to state-of-the-art (SOTA) accuracy~\cite{das2017mixed}, while delivering a $\approx$1.6X improvement in time-to-train (while Intel Knights Mill offers 2x the throughput over FP32 using 4VNNIW). In section~\ref{sec:performance} we will therefore focus on the kernel performance gains from using 4VNNIW of Knights Mill in our kernel library.  

 \subsection{Framework overview}
\begin{figure}[t!]
	\includegraphics[width=\columnwidth]{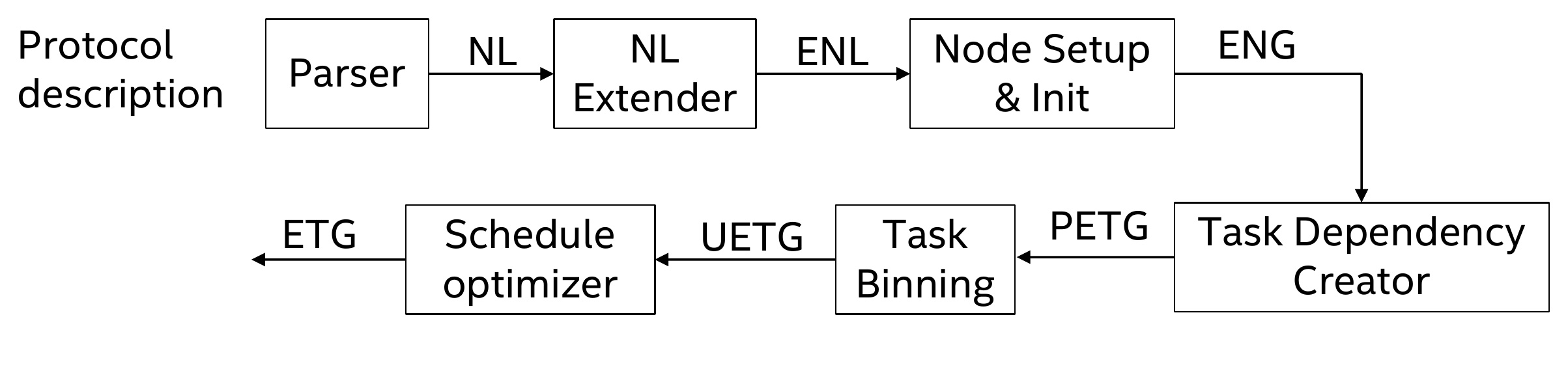}
	\caption{GxM Task Graph Optimization Process}
\label{fig:GxM}
\end{figure}

In this section, we describe our framework for neural network training and inference, called Graph execution Model (GxM). GxM can be seen as very light-weight sibling of Tensorflow~\cite{tensorflow2015-whitepaper}. 
At the core of GxM, is the Execution Task Graph (ETG) that executes the forward, backward propagation and weight gradient update passes for training and only the forward pass for inference. Each node of the ETG is a task that executes one of the three passes when invoked. 
GxM also supports multi-node training -- each ETG node sets up communication end-points in node types that exchange weight gradients, i.e., convolution, batch normalization and inner-product. GxM uses the Intel MLSL library for multi-node training which scales to hundreds of nodes~\cite{mlsl}. 

Figure~\ref{fig:GxM} depicts the flow-chart describing our algorithm to build the ETG. The parser block parses the DNN topology description -- expressed in Protobuf~\cite{protobuf} format -- into a Network List (NL) object. The NL Extender block adds ``Split" nodes -- that perform tensor distribution and reduction in the forward and back propagation steps, respectively -- to the NL, creating the Extended NL (ENL). Then an Extended Node Graph (ENG) is created which is transformed by considering forward and backward dependencies into a Preliminary ETG (PETG). In the next step an Un-optimized ETG (UETG) is created by using a task binning approach and finally duplicates are eliminated which results into the final ETG.

\section{Performance Evaluation}
\label{sec:performance}

\begin{table}
\fontsize{6}{4}\selectfont
 \begin{tabularx}{\columnwidth}{|X| l | l |X|X|X|X| X || X| l | l |X|X|X|X| X |}
 \hline
ID	&	C	&	K	&	H	&	W	&	R	&	S	&	str	&	ID	&	C	&	K	&	H	&	W	&	R	&	S	&	str	\\
 \hline
 \hline
1	&	3	&	64	&  \fontsize{5}{5}\selectfont 224	&\fontsize{5}{5}\selectfont	224	&	7	&	7	&	2	&	11	&	512	&	1024	&	28	&	28	&	1	&	1	&	2	\\
  \hline
2	&	64	&	256	&	56	&	56	&	1	&	1	&	1	&	12	&	512	&	256	&	28	&	28	&	1	&	1	&	2	\\
 \hline
3	&	64	&	64	&	56	&	56	&	1	&	1	&	1	&	13	&	256	&	256	&	14	&	14	&	3	&	3	&	1	\\
 \hline
4	&	64	&	64	&	56	&	56	&	3	&	3	&	1	&	14	&	256	&	1024	&	14	&	14	&	1	&	1	&	1	\\
 \hline
5	&	256	&	64	&	56	&	56	&	1	&	1	&	1	&	15	&	1024	&	256	&	14	&	14	&	1	&	1	&	1	\\
 \hline
6	&	256	&	512	&	56	&	56	&	1	&	1	&	2	&	16	&	1024	&	2048	&	14	&	14	&	1	&	1	&	2	\\
 \hline
7	&	256	&	128	&	56	&	56	&	1	&	1	&	2	&	17	&	1024	&	512	&	14	&	14	&	1	&	1	&	2	\\
 \hline
8	&	128	&	128	&	28	&	28	&	3	&	3	&	1	&	18	&	512	&	512	&	7	&	7	&	3	&	3	&	1	\\
 \hline
9	&	128	&	512	&	28	&	28	&	1	&	1	&	1	&	19	&	512	&	2048	&	7	&	7	&	1	&	1	&	1	\\
 \hline
10	&	512	&	128	&	28	&	28	&	1	&	1	&	1	&	20	&	2048	&	512	&	7	&	7	&	1	&	1	&	1	\\
 \hline
\end{tabularx}
\caption{ResNet-50 layers specifications, on KNM we used a minibatch of 70, on SKX a minibatch of 28.}
\label{tab:resnet_layers}
\end{table}

\begin{figure*}[h!]
\centering
\includegraphics[width=1.7\columnwidth]{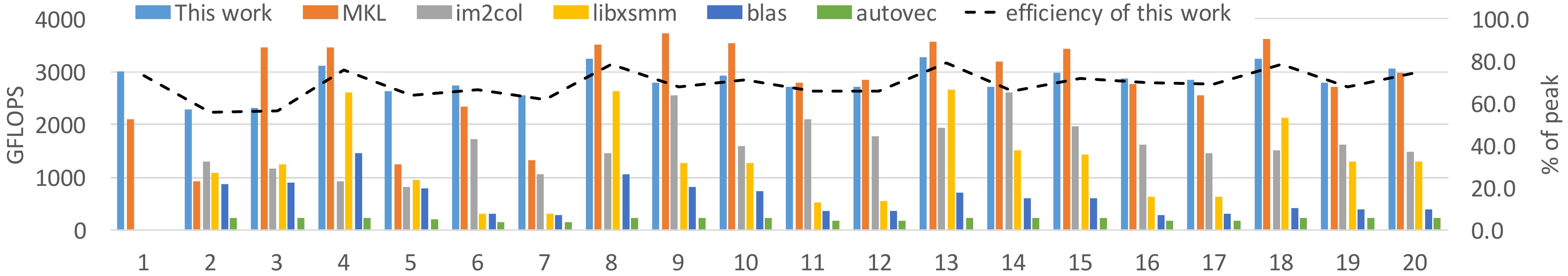}
\vspace{-5pt}
\caption{Performance of ResNet-50 forward propagation on single-socket Skylake (SKX). The x axis corresponds to the layer ids in Table~\ref{tab:resnet_layers}.}
\label{fig:fwd_skx}
\end{figure*}
\begin{figure*}[h!]
\centering
\includegraphics[width=2\columnwidth]{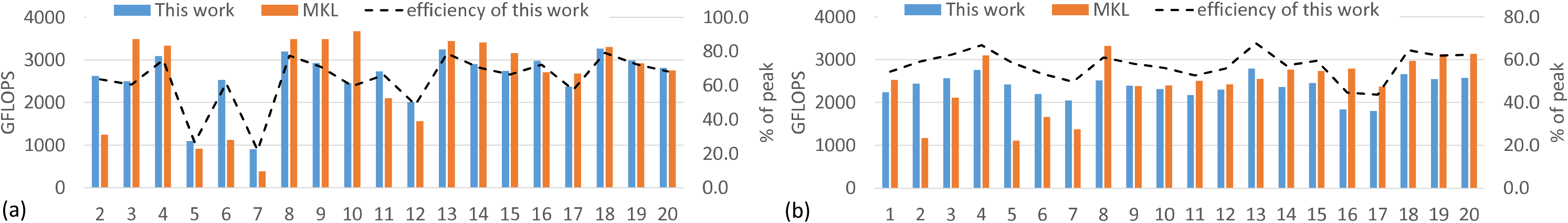}
\vspace{-5pt}
\caption{Performance of ResNet-50 (a) backward propagation and (b) weight update propagation on Skylake (SKX).  The x axes correspond to the layer ids in Table~\ref{tab:resnet_layers}.}
\label{fig:bwd_upd_skx}
\end{figure*}

After the discussion of how the convolution kernels are generated and how they are orchestrated in GxM, we evaluate the performance
on several architectures. All experiments are 
carried out on two similar testbeds, each a 16-node cluster with Intel\textsuperscript{\textregistered} Omnipath interconnect. Each cluster uses its own
48-port Omnipath switch to reduce noise. The following CPU options are used per node:

\noindent \textbf{Skylake-SP (SKX):} 2 Intel\textsuperscript{\textregistered} Scalable  Xeon\textsuperscript{\textregistered} 8180 processors with 28 cores each with 96 GB\,DDR4 2666 main memory at 2.3\,GHz (AVX512) Turbo at 205W TDP. The stream
triad performance of a single socket is 105\,GB/s and one socket reaches 3.8 TFLOPS for SGEMM using AVX512.

\noindent \textbf{Knights Mill (KNM):} Intel\textsuperscript{\textregistered} Xeon Phi\textsuperscript{\texttrademark} 7295 processor with 72 cores and 16\,GB MCDRAM plus 96\,GB DDR4 2400 main memory at 1.6\,GHz Turbo and 320 W thermal design power (TDP). The stream 
triad performance of a single node is roughly 470\,GB/s and the chip achieves 11.5 TFLOPS of SGEMM performance. KNM achieves this high performance by a 4-way-chained FMA instruction (4FMA) which can be used to implement non-transposed GEMM very efficiently.

The performance evaluation is split into two parts, first we evaluate the kernel-only performance and second we discuss the full graph-based execution performance. All the numbers presented are averages over 20 runs
and the run-to-run variation was determined at $\approx$3\% due to a careful setup of our nodes.  
Throughout this section we will present detailed kernel performance results on the state-of-the-art ResNet-50~\cite{he2016deep} topology. We will also briefly present performance results on the modern Inception-v3 topology~\cite{szegedy2016rethinking}.
In addition to our work, we compare several \emph{alternative} implementations for convolution layers:

\noindent \textbf{im2col}: This is the method of performing convolutions popularized by the Caffe~\cite{jia2014caffe} framework. In this method, the input data are flattened and subsequently standard matrix multiplication calls are performed.

\noindent  \textbf{libxsmm}: This method uses the implementation of the direct convolution loops that are properly blocked to accomodate small matrix multiplications as the innermost microkernel. For the innermost small GEMM kernel we use the high performance LIBXSMM library~\cite{Heinecke:2016:LAS:3014904.3015017}.

\noindent  \textbf{blas}: Same implementation as above, but instead of leveraging LIBXSMM we are using MKL GEMM calls (v2017.0.4).

\noindent  \textbf{autovec}: Same implementation as above, but instead of using MKL GEMM calls, we explicitly spell out the small  GEMM as three nested loops and we rely on the compiler to vectorize automatically the loops (compiler version icc v2017.0.4).

\noindent  \textbf{MKL}: For completeness we benchmark the MKL-DNN library v0.12~\cite{mkldnn} which is specialized for direct convolutions. We want to emphasize here that the work presented in this paper is a research project that represents a multi-year effort. We have already shared many insights/techniques presented in this paper with Intel's MKL software team. Not all of these techniques are productized yet, and some are unique to our work e.g. kernel streams for fusion, complicated fused operators, duality for backward propagation to reduce number of code generators, optimized low precision kernels. We compare the basic implementation of our work (i.e.\ without any layer fusion) to the MKL-DNN library which already is a productization of core ideas presented here, i.e.\ these ideas were originated by the authors of this work and are existent in both code bases.


The layers of the ResNet-50 topology are summarized in Table~\ref{tab:resnet_layers}, where each layer is assigned a layer id in the range 1-20 for the remaining paper.
\subsection{Skylake-SP (SKX) performance evaluation}
Figure~\ref{fig:fwd_skx} illustrates the performance of ResNet-50 forward propagation on Skylake (SKX). The x-axis is indexed based on the ResNet-50 layer id. The left y-axis shows achieved performance for each implementation in GFLOPS, while the right y-axis shows the performance of our implementation (``This work") as a \% of the machine peak.

First we observe that the performance of our work for the majority of the layers lies in the regime of 70\%-80\% of the machine peak.
More specifically, layers with $R=1$ and $S=1$ achieve $\approx$70\% of peak since their operational intensity and the input/output tensor reuse is lower compared to the layers with $R=3$ and $S=3$, which achieve $\approx$80\% of peak. Layers 2-3 attain $\approx$55\% of the peak. The reason is as follows: These layers have a small number of input feature maps and as such the input tensor reuse is further limited. Additionally, the spatial dimensions of the output tensors are large meaning that the process of writing the output tensors is characterized by high bandwidth requirements.

Comparing to MKL, we observe speedups in some layers in the range of $1.1\times$-$1.2\times$. However, for the majority of the layers, the two implementations exhibit similar performance; as explained earlier, many techniques presented in this paper are already shared with Intel's MKL software team and are productized in MKL-DNN.

Comparing to the im2col implementation, our work illustrates speedup up to $3\times$, while comparing to the GEMM based approaches (libxsmm,blas) our works yields speedups up to $9\times$ (with the libxsmm based implementation being consistently faster than the ``blas" variant). Finally, the compiler vectorized implementation is by far the slowest, with our work being up to $16\times$ faster. These results highlight the necessity to leverage specialized implementations of direct convolutions, like the one presented in this work, that optimize the data movement, avoid redundant data transformations and leverage the underlying platform's features (e.g.\ cache, vectorized instructions, software prefetching, streaming stores) to the greatest extent. For the backward and weight update propagation we show results only for our work and MKL-DNN.

Figures~\ref{fig:bwd_upd_skx} (a) and (b) show the performance of the backward and weight update propagation passes respectively. The performance of backward propagation is similar to the forward propagation. This behavior is expected since our implementation employs algorithmic duality for backward propagation, as described in Section~\ref{subsec:duality}. Layers with $stride=2$ constitute notable exceptions, where the performance deteriorates. In these cases, the input gradient tensors (the outcome of the convolutions) expand in size (compared to the gradient output tensors) and therefore the corresponding layers exhibit higher write bandwidth requirements. Finally, the efficiency of the weight update propagation kernels is 10\%-15\% lower than the corresponding efficiency of the forward propagation kernels. This degradation is a result of the required weight reduction that is described in Section~\ref{subsec:wup}.

Regarding the performance of the convolution kernels in the Inception-v3 topology, the average performance of our work across all topology's layers is 2833, 2695 and 2621 GFLOPS for the forward, backward and weight propagation passes respectively. The corresponding average performance of the MKL-DNN library is 2758, 2434 and 2301 GFLOPS.

\subsection{Knights Mill (KNM) performance evaluation}
\begin{figure*}[!t]
\centering
\includegraphics[width=1.7\columnwidth]{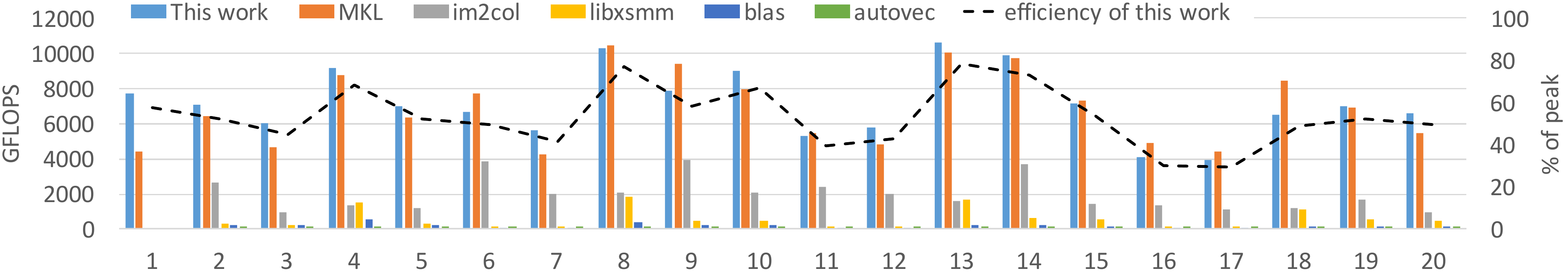}
\vspace{-5pt}
\caption{Performance of ResNet-50 forward propagation on Knights Mill (KNM)}
\label{fig:fwd_knm}
\end{figure*}

Figure~\ref{fig:fwd_knm} illustrates the performance of ResNet-50 forward propagation on Knights Mill (KNM). Layers with $R=1$ and $S=1$ achieve $\approx$55\% of peak since their operational intensity and the input/output tensor reuse is lower compared to the layers with $R=3$ and $S=3$, which achieve 70\%-75\% of peak. The only notable difference compared to the efficiency of the convolutions on the SKX platform pertains to the convolutions with $R=1$ and $S=1$, where on SKX they exhibited efficiency $\approx$70\%. This difference can be justified by considering the roofline models for the KNM and SKX platforms. Each KNM core can attain 54.4 GB/s READ and 27 GB/s WRITE L2 bandwidth, whereas the core's peak performance is 192 GFLOPS. On the other hand, each SKX core can attain 147 GB/s READ and 74 GB/s WRITE L2 bandwidth, whereas the core's peak performance is 147 GFLOPS. Even though the layers with $R=1$ and $S=1$ are properly blocked to maximize cache reuse, their operational intensity lies in the KNM's roofline regime which is characterized as L2 bandwidth bound, whereas for SKX's  roofline model, such operational intensity lies in a regime that is closer to the compute bound region. On the contrary, layers with $R=3$ and $S=3$ have substantially higher operational intensity (e.g. see Section~\ref{subsec:cache_block}) and therefore achieve close to compute peak performance even on KNM.

\begin{figure*}[!t]
\centering
\includegraphics[width=2\columnwidth]{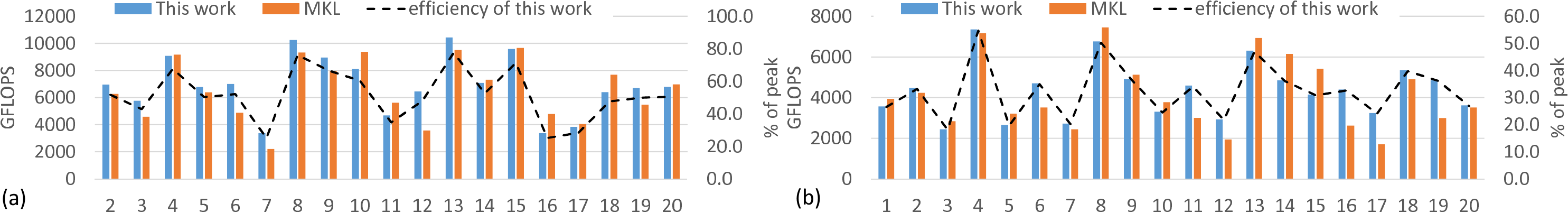}
\vspace{-5pt}
\caption{Performance of ResNet-50 (a) backward propagation and (b) weight update propagation on Knights Mill (KNM)}
\label{fig:bwd_upd_knm}
\end{figure*}

Figures~\ref{fig:bwd_upd_knm} (a) and (b) show the performance of the backward and weight update propagation passes respectively. The performance of backward propagation is similar to the forward propagation. On KNM, the efficiency of the weight update propagation kernels is in the range of 20\%-55\%. There are two reasons behind this behavior. First, the weight reduction overhead discussed in Section~\ref{subsec:wup} is even more emphasized on KNM compared to SKX; KNM does not have a shared Last Level Cache (unlike SKX) that absorbs most of the reduction-involved data movement. Instead, this reduction stresses the memory bandwidth and degrades the overall performance. Second, in order to make use of KNM's 4FMA instruction in the weight gradient update microkernel, we have to transpose upfront the spatial ($W$) and the innermost feature map dimensions of the gradient input tensor; this is a memory bound operation and further degrades the performance of the overall weight update kernel. Regarding the performance of the convolution kernels in the Inception-v3 topology, the average performance of our work across all topology's layers is 6647, 5666 and 4584 GFLOPS for the forward, backward and weight propagation passes respectively. The corresponding average performance of the MKL-DNN library is 7374, 5953 and 4654 GFLOPS.
\begin{figure*}[!t]
\centering
\includegraphics[width=2\columnwidth]{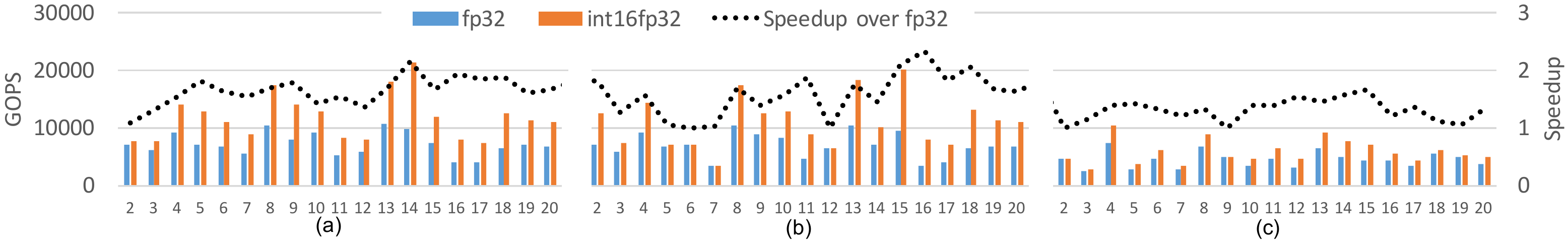}
\vspace{-5pt}
\caption{Performance of ResNet-50 (a) forward propagation, (b) backward propagation and (c) weight update propagation on Knights Mill (KNM) with reduced precision kernels}
\label{fig:LP}
\end{figure*}

Figures~\ref{fig:LP} (a), (b) and (c) show the performance of the ResNet-50 forward, backward and weight update kernels with reduced precision on Knights Mill (KNM) as discussed in Section~\ref{subsec:lp}. For the forward and the backward propagation kernels, the average speedups of the reduced precision kernels over the single precision kernels are $1.63\times$ and $1.58\times$ respectively. There are mainly two reasons that prevent these low precision kernels from achieving $2\times$ speedup. First, even though the reduced precision computation involves tensors with half size compared to the kernels with single precision, the kernel's output is still in 32 bits. Hence, the output related data movement does not show any speedup over the corresponding output data movement of the single precision kernels (i.e.\ they have the same bandwidth requirements). Second, we have to restrict the length of the FMA accumulation chain in the microkernels in order to avoid overflows in the output registers~\cite{das2017mixed}. As a consequence, the restricted accumulation chain limits the register data reuse discussed in Section~\ref{subsec:cache_block} and further decreases the attained speedup. For the weight update reduced precision kernels, the average speedup over the single precision kernels is $1.3\times$. In addition to the two aforementioned reasons, the weight gradient tensors's reduction in this pass also involves tensors with 32-bit values which imposes additional movement of 32-bit data and further diminishes the benefits of the computational speedup.

When comparing our work to MKL-DNN, we recognize that our work is in several cases slower (up to 20\%) than MKL-DNN in Figure~\ref{fig:fwd_skx} (SKX performance) but not in Figure~\ref{fig:fwd_knm} (KNM performance). The reason for this is the used instruction sequence. Our work features an instruction sequence that optimizes across the Xeon and Xeon Phi family processors and aims for strong scaling of deep learning training. This means, our work utilizes AVX512F FMA instructions with \emph{fused memory operand} and uses as few as possible tensor elements for efficient vectorization and parallelization. However, fused memory operands suffer from roughly a 15\% performance hit on Xeon SKX as the instruction is broken down into several micro-ups in the processor's backend. This can be worked-around by using more aggressive blocking over output channels which might result into lower performance when strong scaling the deep learning training tasks as we shuffle simple parallelism from thread level into vector level. However, in our benchmark this is not the case and therefore MKL-DNN is in few cases faster than our work on SKX. On KNM (Figure~\ref{fig:fwd_knm}) the same instruction sequence is used for our work and MKL-DNN, hence the performance is similar.

\subsection{Full Topology Performance}
\label{sec:gxmperformance}
\begin{figure}[t!]
\centering
\includegraphics[width=0.9\columnwidth]{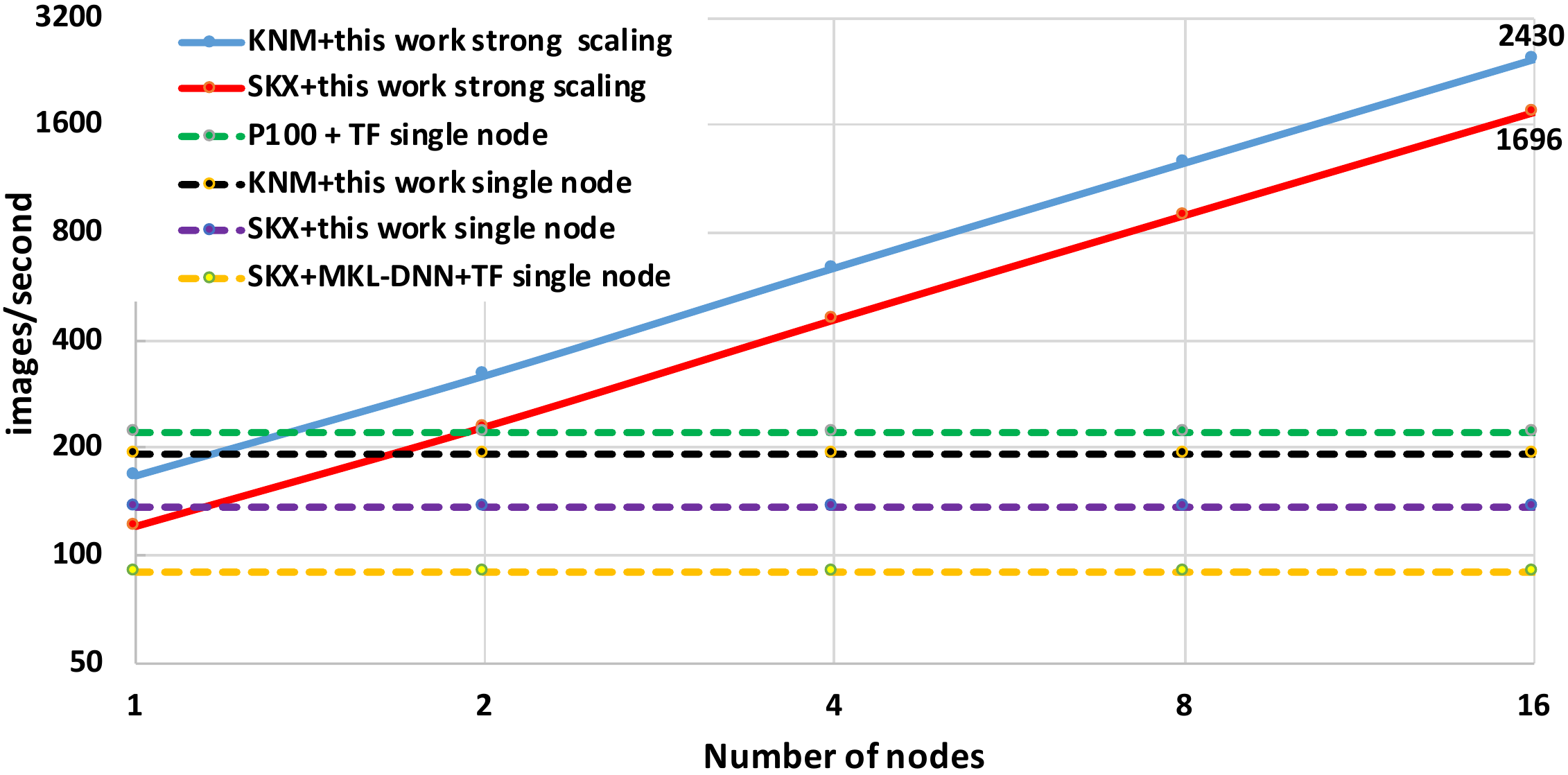}
\caption{End-to-end performance for training of ResNet-50.}
\label{fig:scaling}
\end{figure}

Finally, we evaluate full end-to-end performance of training ResNet-50 and Inception-v3 using our light-weight GxM framework. We compare the obtained performance to Tensorflow-1.6 using MKL-DNN as a kernel library and Tensorflow using cuDNN on a NVidia P100 GPU with performance numbers provided by Google~\cite{googlebench}. Additionally, we strong-scale to the full 16 nodes (896 SKX cores and 1152 KNM cores) of our testbed to demonstrate that efficient deep learning training does not end at the coherent memory boundary. The experiment was carried out using single precision as SKX donesn't have efficient low precision support. The performance summary is provided in Figure~\ref{fig:scaling}. In case of multinode training we only use 62 cores per KNM for compute as 8 cores are used for driving the network via the MLSL library~\cite{mlsl}. On SKX we have to set 4 cores aside for communication which leaves us with 52 compute-cores per node. As shown in Figure~\ref{fig:scaling}, this setting allows us to achieve $\approx$ 90\% parallel efficiency (hence we skiped an ideal scaling line in the plot) on both systems when comparing 1 to 16 nodes's performance. In total we were able to obtain 2430 img/s training performance on 16 nodes of KNM and 1696 img/s on 16 nodes of SKX. This excellent scaling is achieved by using data-parallelism\cite{bennun2018demyst}. The allreduce of the gradient weights in the backward pass is completely overlapped by using MLSL. At single node level, our implementations achieves 192 img/s on a KNM and 136 img/s on a dual-socket SKX node. For comparison, a single NVidia P100 GPU achieves in FP32 219 img/s~\cite{googlebench} and Tensorflow+MKL-DNN was measured at 90 img/s for dual-socket SKX~\cite{mkltfbench}. We also see that the framework can add a huge performance tax. In previous sections we concluded that our presented approach and MKL-DNN achieve comparable kernel performance, but most of this good MKL-DNN's performance is lost during framework integration (Tensorflow in this case) for various reasons such as the lack of fusion, inefficient scratch memory allocation or thread scheduling, to name just a few. End-to-end our work achieves a roughly 2$\times$ speed-up while still converging to the same SOTA accuracies, e.g. 74.5\% Top-1 accuracy for ResNet-50. Additionally, we executed Inception-v3 and the obtained single node numbers confirm the ResNet-50 picture: our solution was measured at 98 img/s for KNM and 84 img/s for SKX. Tensorflow+MKL-DNN achieved 58 img/s on SKX~\cite{mkltfbench} whereas Tensorflow+cuDNN was timed at 142 img/s on NVidia P100~\cite{googlebench}. These results show that CPU can offer competitive time-to-train for (distributed) deep learning training applications, while scaling similar as GPU-based architectures using MLSL-like techniques \cite{horovod}.

\section{Conclusions}
\label{sec:conclusion}

In this work we derived and defined the ingredients of fast direct convolution kernels for training CNNs on modern CPU architectures. This was done by  demonstrating how JIT compilation can be leveraged to obtain a streamlined code which runs a perfectly-chained sequence of small GEMM operations. Additionally, we provided insights on how the combinatorial explosion in the number of required kernels due to layer fusion in deep neural nets can be handled by our approach. We presented a two-step performance assessment: first we evaluated the kernel efficiency for various topologies and second we presented the end-to-end fully-integrated CNN training performance. At kernel level we were able to achieve up to 80\% of  peak performance and end-to-end we were able to outperform optimized Tensorflow  implementations by 1.5$\times$-2.3$\times$. This proves that CPUs can be a competitive alternative when training neural nets. Last but not least, we strong-scaled our framework to $\approx$1000 cores with $\approx$ 90\% parallel efficiency.
\FloatBarrier
\IEEEtriggeratref{13}

\bibliographystyle{IEEEtran}
\clearpage
\bibliography{IEEEabrv,references}

\scriptsize
\noindent Optimization Notice: Software and workloads used in
performance tests may have been optimized for performance only on
Intel microprocessors.  Performance tests, such as SYSmark and
MobileMark, are measured using specific computer systems,
components, software, operations and functions.  Any change to any
of those factors may cause the results to vary.  You should
consult other information and performance tests to assist you in
fully evaluating your contemplated purchases, including the
performance of that product when combined with other products.
For more information go to http://www.intel.com/performance.

\noindent Intel, Xeon, and Intel Xeon Phi are trademarks of Intel Corporation in the U.S. and/or other
countries.

\normalsize

\cleardoublepage

\section{Artifact Description Appendix: Anatomy Of High-Performance Deep Learning Convolutions On SIMD Architectures}

\subsection{Abstract}

This artifact description sketches how to obtain the several software packages needed, how they are compiled and how the reported performance can 
be re-measured.

\subsection{Description}

\subsubsection{Check-list (artifact meta information)}

{\small
\begin{itemize}
  \item {\bf Algorithm: } direct convolutions for deep learning training
  \item {\bf Program: } This work is available via github under BSD (\url{https://github.com/hfp/libxsmm}), MKL-DNN (\url{https://github.com/intel/mkl-dnn}) version v0.12
  \item {\bf Compilation: } make and cmake
  \item {\bf Data set: } synthetic for kernel tests, imagenet 1.2M for full training \url{http://www.image-net.org/}
  \item {\bf Run-time environment: } Linux
  \item {\bf Hardware: } Intel Xeon Scalable Processor (Skylake), Intel Xeon Phi (Knights Mill), high performance interconnect is recommended
  \item {\bf Execution: } Via shell scripts/job scheduler
  \item {\bf Output: } timings and accuracies from logfiles, dumped weights in case of full topology training which can be used for inference tasks afterwards
  \item {\bf Experiment workflow: } see below
  \item {\bf Experiment customization: } different dataset for training can be chosen, different hardware platforms can be used.
  \item {\bf Publicly available?: } yes, on github, BSD license
\end{itemize}
}

\subsubsection{How software can be obtained (if available)}

Via github (\url{https://github.com/hfp/libxsmm}). 

\subsubsection{Hardware dependencies}

Intel Xeon Scalable Processor (Skylake), Intel Xeon Phi (Knights Mill) for the results in this paper. The kernel JITer presented here, also supports Intel SSE3, Intel AVX, Intel AVX2 platforms which is literally every x86 CPU since 2006.

\subsubsection{Software dependencies}

\begin{itemize}
  \item 64-bit Linux or Mac-OS. 32-bit OS is not supported. 
  \item GCC, Clang, PGI, Intel or Cray C/C++ compiler
  \item MPI library
  \item OpenCV
  \item Protobuf
  \item boost
  \item LMDB
  \item a BLAS library for fallback code paths
\end{itemize}

\subsubsection{Datasets}

All layer performance runs presented in this work were carried out with runs which auto generate input data. For ResNet-50/Inception-v3 based Imagenet training, the imagenet dataset needs to be provided through a LMDB database.

\subsection{Installation}

\subsubsection{This Software}

\begin{scriptsize}
\begin{verbatim}
git clone https://github.com/hfp/libxsmm.git
cd libxsmm
make realclean && make AVX=3 OMP=1 STATIC=1
cd samples/deeplearning/cnnlayer
make realclean && make AVX=3 OMP=1 STATIC=1
\end{verbatim}
\end{scriptsize}

For running GxM, please refer to our github page as several scripts need to be adjusted, dependencies need to be built from source (see list), etc. This page already
has a detailed description of what is needed here.

\subsubsection{MKL-DNN}

Please follow the latest instruction for running \texttt{benchdnn} on the wikipage of MKL-DNN.

\subsection{Experiment workflow}

\subsubsection{This Software}

Then we can run ResNet-50 and Incpetion-v3 layers on single-socket Skylake
\begin{scriptsize}
\begin{verbatim}
export OMP_NUM_THREADS=28
export KMP_AFFINITY=granularity=fine,compact,1,0
./run_resnet50.sh 28 1000 1 f32 F L 1
./run_resnet50.sh 28 1000 1 f32 B L 1
./run_resnet50.sh 28 1000 1 f32 U L 1
./run_googlenetv3.sh 28 1000 1 f32 F L 1
./run_googlenetv3.sh 28 1000 1 f32 B L 1
./run_googlenetv3.sh 28 1000 1 f32 U L 1
\end{verbatim}
\end{scriptsize}
and on Knights Mill
\begin{scriptsize}
\begin{verbatim}
export OMP_NUM_THREADS=70
export KMP_AFFINITY=granularity=fine,compact,1,2
./run_resnet50.sh 70 1000 1 f32 F L 1
./run_resnet50.sh 70 1000 1 f32 B L 1
./run_resnet50.sh 70 1000 1 f32 U L 1
./run_resnet50.sh 70 1000 1 qi16f32 F L 1
./run_resnet50.sh 70 1000 1 qi16f32 B L 1
./run_resnet50.sh 70 1000 1 qi16f32 U L 1
./run_googlenetv3.sh 70 1000 1 f32 F L 1
./run_googlenetv3.sh 70 1000 1 f32 B L 1
./run_googlenetv3.sh 70 1000 1 f32 U L 1
\end{verbatim}
\end{scriptsize}

\subsubsection{MKL-DNN}

Please follow the latest instruction for running \texttt{benchdnn} on the wikipage of MKL-DNN.

\subsection{Evaluation and expected result}

Performance can be simply evaluated by console output provided by our simple layer benchmark in GFLOPS and runtime in ms. The GxM framework reports time per iteration and img/s as console output as well, the most important performance figures in case of CNN training.

Numerical accuracy is provided also by both tests: the layer example runs a simple loop nest as reference code for each convolution operation. The JIT is compared using several norms (Linf of absolute error, L2 of absolute error, Linf of relative error, L2 of relative error). In case of the light-weight graph execution model, after each iteration the current loss is reported and after each epoch the current Top-1 and Top-5 accuracies on the currently trained neural net are reported.

\subsection{Experiment customization}

Whereas the provided scripts focus on the architectures covered in this paper, both, our simple layer benchmark as well as the GxM framework can be easily compiled and run on many different x86 platforms not limited to Intel processors. Based on the information provided here and our github page, it should be fairly simple for the user to adjust the parameters in the provided run scripts.

\subsection{Notes}

Our github pages contains far more information than covered here, e.g. on debugger support,performance profile tools support of out JITer and custom configuration of our library.

\end{document}